\documentclass[aps,prd,twocolumn,nofootinbib,floatfix,showpacs,superscriptaddress]{revtex4-1}
\usepackage{amsmath,amsfonts,amssymb,bm}
\usepackage[utf8]{inputenc}
\usepackage{graphicx}
\usepackage{color}
\usepackage{subfigure}
\usepackage{multirow}
\usepackage{textcomp}
\usepackage{slashed}

\definecolor{purple}{rgb}{0.5,0,0.5}
\definecolor{blue}{rgb}{0.0,0,1.0}

\usepackage[colorlinks=true, pdfstartview=FitV, linkcolor=purple, citecolor= purple, urlcolor=blue]{hyperref}
\begin{document}


\title{Linking continuum and lattice quark mass functions via an effective charge}

\author{Lei Chang}
\affiliation{School of Physics, Nankai University, Tianjin 300071, China}

\author{Yu-Bin Liu}
\affiliation{School of Physics, Nankai University, Tianjin 300071, China}

\author{Kh\'epani Raya}
\affiliation{School of Physics, Nankai University, Tianjin 300071, China}
\affiliation{Instituto de Ciencias Nucleares, Universidad Nacional Autónoma de México, Apartado Postal 70-543, C.P. 04510, CDMX, México }


\author{J. Rodr\'{\i}guez-Quintero}
\affiliation{Department of Integrated Sciences and Center for Advanced Studies in Physics, Mathematics and Computation,
University of Huelva, E-21071 Huelva, Spain}

\author{Yi-Bo Yang}
\affiliation{hCAS Key Laboratory of Theoretical Physics, Institute of Theoretical Physics, Chinese Academy of Sciences, Beijing 100190, China}
\affiliation{School of Fundamental Physics and Mathematical Sciences, Hangzhou Institute for Advanced Study, UCAS, Hangzhou 310024, China}
\affiliation{International Centre for Theoretical Physics Asia-Pacific, Beijing/Hangzhou, China}

\date{\today}

\begin{abstract}

The quark mass function is computed both by solving the quark propagator Dyson-Schwinger equation and from lattice simulations implementing overlap and Domain-Wall fermion actions for valence and sea quarks, respectively. The results are confronted and seen to produce a very congruent picture, showing a remarkable agreement for the explored range of current-quark masses. The effective running-interaction is based on a process-independent charge rooted on a particular truncation of the Dyson-Schwinger equations in the gauge sector, establishing thus a link from there to the quark sector and inspiring a correlation between the emergence of gluon and hadron masses.  

\end{abstract}

%

\maketitle

\section{Introduction}

The non-Abelian nature of QCD leads to fascinating consequences in nuclear and hadronic physics, such as quark-gluon confinement and the emergence of hadron masses~\cite{Roberts:2020hiw,Roberts:2021nhw}. Naturally, this is thightly connected with the way that QCD's fundamental degrees of freedom, quarks and gluons, interact. At its most fundamental level, this needs to be understood from QCD's Green functions, the gauge sector of the theory playing an essential role in the strong interaction mechanism. The nonperturbative self-interacting nature of gluons is not only responsible for the ultraviolet (UV) asymptotic freedom but, presumably, has to be also related to the infrared (IR) slavery of colored objects. Precisely within the IR domain, properties of the low-dimension gluon Green's functions have been argued to entail profound dynamical implications. In particular, longitudinally coupled massless poles comprised in the nonpertubative three-gluon vertex function have been shown to trigger a dynamical mass generation mechanism for the gluon~\cite{Cornwall:1981zr,Cornwall:1989gv,Binosi:2009qm,Ibanez:2012zk}. This accounts for the observed saturation of the gluon two-point Green's function at vanishing momentum~\cite{Cucchieri:2007md,Boucaud:2008ky,Fischer:2008uz,Dudal:2008sp,Bogolubsky:2009dc,Oliveira:2009eh,Tissier:2010ts,Ayala:2012pb}; while logarithmic singularities in both the kinetic term of the two-point function and non-transverse structures of the three-gluon vertex appear to be intimately connected, and related to properties of the ghost two-point function~\cite{Aguilar:2013vaa,Athenodorou:2016oyh}. 

All these features can be conveniently accounted by solving the relevant QCD Dyson-Schwinger equations (DSEs), truncated with a scheme based on a combination of the pinch technique\,\cite{Cornwall:1981zr,Cornwall:1989gv,Pilaftsis:1996fh,Binosi:2009qm} and background field method\,\cite{Abbott:1981ke} (PT-BFM). Within this framework, a bridge from QCD gauge to matter sectors have been recently paved~\cite{Binosi:2014aea}, linking therefore the emergence of gluon and hadron masses~\cite{Roberts:2021nhw}. The connection roots on a sensible definition of the running-interaction for the quark propagator DSE, widely dubbed as gap equation, the interaction resulting from a combination of continuum and lattice QCD (lQCD) analyses of QCD's gauge sector~\cite{Aguilar:2009nf}. Elaborating further on this, a renormalization-group-invariant (RGI) process-independent (PI) effective charge of QCD has been derived from this running-interaction in Refs.~\cite{Binosi:2016nme,Rodriguez-Quintero:2018wma}; and subsequently refined by employing modern lQCD configurations~\cite{Cui:2019dwv}.

Additionally, an alternative effective charge phenomenologically defined to drive the all-orders DGLAP evolution of pion distribution functions have been seen to agree with the PI charge within the IR domain, and make then contact smoothly with the UV well-known perturbative behaviour defined by the evolution kernel~\cite{Cui:2020dlm,Cui:2020tdf}. In this paper, we make one further step by redefining the running-interaction from this last phenomenological effective charge, then apply it to solve the gap equation and derive therefrom the quark mass function\footnote{The reader is referred to \cite{Roberts:2020hiw,Huber:2018ned,Fischer:2018sdj} for recent reports in which the modern formulation of the gap equation is discussed}. We also compute this mass function from lQCD gauge configurations with three dynamical Domain-Wall fermions~\cite{Blum:2014tka,Mawhinney:2019cuc}, tuning the current quark mass by the use of different lattice set-up's, and compare to the corresponding gap-equation solutions. We also solve the gap equation in the chiral limit and, evaluating then the pion decay constant and the quark condensate in this limit, verify the Gell-Mann-Oakes-Renner formula~\cite{GellMann:1968rz}. 

\section{The gap equation and the effective charge}
\label{sec:gapEq}

The quark propagator is, typically, the first basic ingredient for any hadron physics study based upon continuum functional methods and, more specifically, within the DSEs framework~\cite{Roberts:1994dr}. Let us write 
\begin{equation}\label{eq:quark-prop0}
S_{f(0)}(p) =\left(i \gamma \cdot p + m_f^{\textrm{bm}}\right)^{-1} \;, 
\end{equation}
for the bare tree-level quark propagator of flavor $f$, where $ m_f^{\textrm{bm}}$ stands for its bare-quark mass. 
This propagator is to be then \emph{dressed} by incorporating all possible QCD quantum corrections and, subsequently, regularisation and renormalisation prescriptions need to be implemented; the latter introducing a renormalization point $\zeta$. A $\zeta$-dependent current-quark mass $m_f^\zeta$ will thus result, being directly related to the bare-quark mass via Slavnov-Taylor identities~\cite{Slavnov:1972fg,Taylor:1971ff} (STI). The fully dressed propagator can be obtained by solving the gap equation, 
\begin{subequations}
\label{eq:gapEqs}
\begin{eqnarray} 
\label{eq:quark-prop1}
S_f^{-1}(p)&=&  Z_f^{-1}(p^2) (i \gamma \cdot p + M_f(p^2)) \\ 
\label{eq:gapEq1}
&=& Z_2  S_{f(0)}^{-1}(p) + \Sigma_f(p) \;,\\
\Sigma_f(p) &=& \frac{4}{3}Z_1 \int_{dq}^{\Lambda} g^2 D_{\mu\nu}(p-q)  \gamma_\mu S_f(q) \Gamma_\nu^{f}(p,q)\;. 
\label{eq:selfE}
\end{eqnarray}
\end{subequations}
In Eq.\,\eqref{eq:quark-prop1}, the dressed quark propagator appears recast, keeping the analogy with its bare counterpart given by Eq.\,\eqref{eq:quark-prop0}, in terms of the dressing functions $Z_f(p^2)$ and $M_f(p^2)$, which capture both the perturbative and nonperturbative facets of the propagator. In particular, the latter one, independent of $\zeta$, corresponds to the constituent quark mass function which we shall focus on in this work. Eqs.\,\eqref{eq:gapEq1} and\,\eqref{eq:selfE} properly display the gap equation, where $\int_{dq}^\Lambda = \int^\Lambda \frac{d^4q}{(2\pi)^4}$ stands  for  a  Poincar\'e  invariant  regularized  integration,  with  $\Lambda$  for the regularization scale. The rest of the pieces carry their usual meaning: $D_{\mu\nu}$ is the gluon propagator and $\Gamma_\nu$ the fully-dressed quark-gluon vertex (QGV); $Z_{1,2}$ are the QGV and quark wave-function renormalization constants, respectively; and $g$ is the Lagrangian coupling constant. Every piece in Eqs.\,\eqref{eq:gapEqs} depends on $\zeta$, although the explicit dependence has been omitted for simplicity. \color{black} Each Green's function involved obeys its own DSE, thus forming an infinite tower of coupled integral equations. 

The derivation of tractable solutions from this infinite set of equations requires a truncation scheme, conveniently grounded on a certain number of physically sound and mathematically reliable assumptions for some suitable set of Green functions~\cite{Binosi:2016rxz}.  Particularly, in truncating Eq.~\eqref{eq:quark-prop1}, it is typical to assume a particular form for the QGV~\cite{Qin:2020jig,Aguilar:2018epe,Serna:2018dwk,Albino:2018ncl,Bashir:2011dp}, supplemented with a suitable choice for the QCD's interaction strength~\cite{Qin:2011dd,Qin:2020jig} assumed to incorporate effectively the QCD dynamics and accounting thus for hadron observables. Additionally, several requirements should be imposed to the QGV, as \emph{e.g.} gauge invariance, multiplicative renormalizability or absence of kinematic singularities~\cite{Bermudez:2017bpx}.

Owing to a veracious expression of key dynamical QCD features in the gauge sector, as the infrared saturation of gluon propagators and the massless nature of ghost propagators\,\cite{Cucchieri:2007md,Bogolubsky:2007ud,Bogolubsky:2009dc,Oliveira:2009eh,Ayala:2012pb,Aguilar:2004sw,Aguilar:2006gr,Aguilar:2008xm,Boucaud:2008ky,Boucaud:2008ji,Fischer:2008uz,Dudal:2008sp,RodriguezQuintero:2010wy,Tissier:2010ts}, the gap-equation quark-antiquark scattering kernel has been recently approached from this gauge sector and shown to be consistent with a matter-sector construction of the same kernel\,\cite{Binosi:2014aea}, based on a nonperturtative symmetry-preserving truncation of the bound-state equations and further comparison with empirical data\,\cite{Chang:2009zb,Chang:2010hb,Chang:2013pq}. Within this framework, Eq.\,\eqref{eq:selfE} can be rewritten as (after fixing the Landau gauge)
\begin{eqnarray}\label{eq:quark-prop2}
&&\Sigma_f(p)= \frac{4}{3} Z_2 \int_{dq}^\Lambda 4\pi \widehat{d}(k^2)T_{\mu\nu}(k) \gamma_\mu S_f(q) \widehat{\Gamma}^f_\nu(p,q)\;,\\
&&T_{\mu\nu}(k)=\delta_{\mu\nu}-k_\mu k_\nu / k^2,\;k=p-q \;; \nonumber
\end{eqnarray}
where one capitalizes on the PT-BFM scheme, which makes possible a convenient redefinition of the QCD Green's functions, \emph{via} a systematical rearranging of classes of diagrams in their DSEs, such that they result to satisfy linear STIs. Rooting on the latter: 
\\

\emph{(i) In the gauge sector}, the PT-BFM gluon vacuum polarisation captures the required renormalization-group (RG) logarithmic behavior\,\cite{Aguilar:2009nf}, leading therefrom to define a unique QCD running coupling from the gauge-field two-point Green's functions\,\cite{Binosi:2016nme,Rodriguez-Quintero:2018wma,Cui:2019dwv}, 
\begin{subequations}
\begin{eqnarray}
g^2 D_{\mu\nu}(k) \to g^2 \widehat{D}_{\mu\nu}(k) = 4 \pi \widehat{d}(k^2) T_{\mu\nu}(k)\;, \\ \label{eq:dhat}
k^2 \widehat{d}(k^2)=\frac{\alpha_{\text{T}}(k^2)}{[1-L(k^2;\zeta^2)F(k^2;\zeta^2)]^2}\;; 
\end{eqnarray}
\end{subequations}
where $L(k^2;\zeta^2)$ is a longitudinal piece of the gluon-ghost scattering kernel obeying its own DSE\,\cite{Aguilar:2009nf} and $\alpha_{\text{T}}(k^2)$ the running coupling derived from the ghost-gluon vertex `Taylor coupling'~\cite{Sternbeck:2007br,Boucaud:2008gn,Blossier:2012ef}, whose renormalization flow is defined in terms of gluon and ghost dressing functions\footnote{The dressing function is a widely used denomination for the nonperturbative piece of the two-point scalar form factor; \emph{e.g.}, in the case of the gluon propagator: $D^{ab}_{\mu\nu}(k)=\delta^{ab} D_{\mu\nu}(k) = \delta^{ab} T_{\mu\nu}(k) G(k^2)/k^2$}, respectively $G(k^2;\zeta^2)$ and $F(k^2;\zeta^2)$, 
\begin{equation}
\alpha_{\text{T}}(k^2) = \alpha(\zeta^2) G(k^2;\zeta^2)F^2(k^2;\zeta^2)\;;
\end{equation}
with the Lagrangian coupling at the renormalization scale, $\alpha(\zeta^2)=g^2(\zeta^2)/(4\pi)$, as the starting point for the evolution. 
In Eq.\,\eqref{eq:dhat}, the explicit point-renormalization dependence of the dressing functions has been restored to highlight the renormalization-group-independent (RGI) character of $\hat{d}(k^2)$, which naturally enters as the effective running-interaction in the gap-equation quark-antiquark scattering kernel, Eq.\,\eqref{eq:quark-prop2}.  
As made apparent in Eq.\,\eqref{eq:dhat}, one can construct with this running-interaction a quantity endowed with all the UV RG-features of a QCD running coupling\footnote{One also needs\,\cite{Binosi:2016xxu,Cui:2019dwv}: 
$L(k^2,\zeta^2) F(k^2,\zeta^2) \simeq 3 \alpha_T(k^2)/[8\pi]$ at large momenta. 
}. Capitalizing upon the latter, a process-independent (PI) QCD effective charge has been derived in Refs.\,\cite{Binosi:2016nme,Rodriguez-Quintero:2018wma,Cui:2019dwv}, defined through
\begin{equation}\label{eq:dhatfromalpha}
\widehat{d}(k^2) = \widehat{\alpha}(k^2) {\mathcal D}(k^2) \;,
\end{equation}
where ${\mathcal D}(k^2)$ is also a RGI function behaving in both the far-infrared and -ultraviolet as the propagator of a free massive boson and, as explained in Ref.\,\cite{Cui:2019dwv}, obtained from modern lattice QCD (lQCD) estimates of the gluon propagator.   
\\

\emph{(ii) In the matter sector}, the fully-dressed QGV is modified to obey a linear STI, 
\begin{equation}
Z_1 \Gamma^f_\nu(p,q) \to Z_2 \hat{\Gamma}^f_\nu(p,q) \;,
\end{equation}
for which a sensible Ansatz is
\begin{eqnarray}
\label{eq:vertex}
\hat{\Gamma}^f_\nu(p,q)=\Gamma_{\nu}^{f,\text{BC}}(p,q) + \Gamma_{\nu}^{f,\text{ACM}}(p,q) \;.
\end{eqnarray}
The first piece, $\Gamma_{\nu}^{f,\text{BC}}(p,q)$, corresponds to the well known Ball-Chiu vertex~\cite{Ball:1980ay}, which completely determines the longitudinal part of the vertex by the requirement of gauge invariance~\cite{Ward:1950xp,Green:1953te,Takahashi:1957xn}. The second piece is associated with the Anomalous Chromomagnetic Moment (ACM) term~\cite{Chang:2010hb,Binosi:2016wcx}, whose explicit structure can be conveniently written as
\begin{eqnarray}\label{eq:ACMvertex}
&&	\Gamma^{f,\text{ACM}}_\nu(p,q) = \eta\; \sigma_{\nu\alpha}k_\alpha \frac{B_f(p^2)-B_f(q^2)}{p^2-q^2} \mathcal{H}\left(k^2\right)\;;
\end{eqnarray}
still with $k=p-q$; and here $B_f(s)=M_f(s)/Z_f(s)$ and the profile function, $(s/m_0^2)\mathcal{H}(s)=(1-e^{-s/m_0^2})$,  defined such that it controls the ultraviolet convergence and restricts the ACM effects to the infrared domain~\cite{Chang:2010hb}; $m_0$  and $\eta$ being flavor-independent parameters that will be herein tuned by comparison with lQCD data.
\\

Then, replacing Eq.\,\eqref{eq:selfE} with\,\eqref{eq:quark-prop2}, the gap equation can be solved with the running-interaction $\widehat{d}(k^2)$ as a key ingredient for the kernel, thus featuring a very appealing connection between the QCD effective coupling defining the running-interaction and the quark mass function $M_f(q^2)$. Furthermore, a remarkable outcome from applying the ansatz given by Eqs.\,\eqref{eq:vertex} and \eqref{eq:ACMvertex} for the QGV is that all flavor dependence in the resulting quark mass function stems from the choice for the current quark mass, with no further tuning of additional parameters related to the kernel. This \emph{universality} of our bridging from the QCD effective coupling to the quark mass function can and will be herein tested by a comparison of lQCD and gap-equation results obtained with several different current quark masses in the light sector.

\section{Effective Charge and QCD evolution of the pion PDF}\label{sec:DGLAP}
\label{sec:eff-charge}

According to the seminal work of Ref.\,\cite{Grunberg:1982fw}, an alternative, process-dependent approach to determining an ``effective charge'' consists in its definition as being completely fixed by the leading-order term of the canonical perturbative expansion of a given observable. An example of charge thus defined is that fixed by the Bjorken sum rule\,\cite{Bjorken:1966jh,Bjorken:1969mm}, which is compared to the PI charge in Ref.\,\cite{Binosi:2016nme,Cui:2019dwv}. Following the same process-dependent approach, another effective charge, $\widetilde{\alpha}(k^2)$, for which evolution of all pion PDFs moments is completely defined by the one-loop formula, is introduced in Refs.\,\cite{Cui:2019dwv}; and it is therein conjectured to agree within the IR with $\widehat{\alpha}(k^2)$. 

Further on the same track, it has been defined\,\cite{Cui:2020dlm,Cui:2020tdf} 
\begin{eqnarray}
\label{eq:alphaPIparam}
\widetilde{\alpha}(k^2)=\frac{\gamma_m \pi}{\ln \left[\frac{\mathcal{K}^2(k^2)}{\Lambda^2_{\text{QCD}}} \right]}\;,
\end{eqnarray}
where $n_f$ accounts for the number of active quark flavors (within the UV domain) and $\gamma_m=4/\beta_0$, $\beta_0=11-(2/3)n_f$; and the interpolation function, 
\begin{equation}\label{eq:K2}
 \mathcal{K}^2(y)=\frac{a_0^2+a_1y+y^2}{b_0+y}
\end{equation}
with $\{a_0,\;a_1,\;b_0\}=\{0.104(1),0.0975,0.121(1)\}$ (in appropriate powers of GeV$^2$), ensuring a smooth connection between both the correct IR and UV behaviors. For the latter, contact with the 1-loop level perturbation theory is made, with $n_f$=4 and $\Lambda_{\text{QCD}}$=0.234 GeV; while agreement within the IR domain has been required\footnote{For this, we have imposed an accurate parametrization of $\widehat{\alpha}(k^2)$ from Ref.\,\cite{Cui:2019dwv} with Eqs.\,(\ref{eq:alphaPIparam},\ref{eq:K2}) with $\Lambda_T$=0.52 GeV (the MOM-scheme value suggested by Eq.\,\eqref{eq:dhat} for the large-momentum behavior of the PI effective coupling) and then rescaled the so-obtained parameters: $\{a'_0,a'_1,b'_0\} \to \{a_0,a_1,b_1\}=\{a'_0,a'_1,b'_0\} \times (\Lambda_{\text{QCD}}/\Lambda_T)^2$. 
}
\,for $\widetilde{\alpha}(k^2)$ and the most refined estimate of $\widehat{\alpha}(k^2)$, obtained with lQCD propagators resulting from modern gauge-field configurations generated with three domain-wall fermions at the physical pion mass~\cite{Cui:2019dwv}. 

The effective charge represented by Eqs.\,(\ref{eq:alphaPIparam}, \ref{eq:K2}) has been used in Ref.\,\cite{Cui:2020dlm,Cui:2020tdf} to evolve singlet and non-singlet pion parton distribution functions (PDFs) from a given low-momentum hadronic scale $\zeta_H$, at which they can be estimated from valence-quark distributions obtained with continuum functional methods\,\cite{Ding:2019lwe,Ding:2019qlr}, up to a large-momentum experimental scale $\zeta_{\text{ex}}$. Glue, sea- and valence-quark pion PDFs at $\zeta_{\text{ex}}$ come out with QCD evolution\,\cite{Gluck:1999xe}. All-orders evolution with the 1-loop DGLAP kernel supplemented with the effective charge $\widetilde{\alpha}(k^2)$ is assumed and the results are shown to be in excellent agreement with experiments\,\cite{Cui:2020dlm,Cui:2020tdf}. For the sake of a single illustration, the following results ($n_f$=4), 
\begin{subequations}
\label{eq:xs}
\begin{eqnarray}
\langle 2 x(\zeta_{\text{ex}}) \rangle_q &=&   \exp{\left(- \frac 8 {9\pi} S\left(\zeta_H,\zeta_{\text{ex}}\right)\right)} \,,  \\
\langle x(\zeta_{\text{ex}}) \rangle_{\text{sea}} &=& \frac 3 7 + \frac 4 7 \langle 2 x(\zeta_{\text{ex}})  \rangle_q^{7/4} - \langle 2 x(\zeta_{\text{ex}}) \rangle_q \,, \\
\langle x(\zeta_{\text{ex}}) \rangle_{\text{glue}} &=& \frac 4 7 \left( 1 -   \langle 2 x(\zeta_{\text{ex}})  \rangle_q^{7/4} \right) \;,
\end{eqnarray}
\end{subequations}
can be readily derived and shown to display a closed algebraic relation between $\langle x(\zeta_{\text{ex}}) \rangle_q$, $\langle x(\zeta_{\text{ex}}) \rangle_{\text{sea}}$ and $\langle x(\zeta_{\text{ex}}) \rangle_{\text{glue}}$,  the momentum fraction averages for the pion valence-quark ($q$=$u,d$), sea-quarks and glue at the evolved scale; with
\begin{equation}
S\left(\zeta_H,\zeta_{\text{ex}}\right) = \int_{t(\zeta_H)}^{t(\zeta_{\text{ex}})} dt(\zeta) \, \widetilde{\alpha}(t(\zeta)) 
\end{equation}
and $t(\zeta)=\ln{(\zeta^2/\Lambda_{\text{QCD}}^2)}$. Eqs.\,\eqref{eq:xs} only rely on the all-orders QCD evolution from the hadronic scale with the effective charge and can provide with a fairly good description of experimental\,\cite{Novikov:2020snp} and lattice\,\cite{Sufian:2020vzb} results. In addition, they make apparent the momentum sum rule and the limit of glue and sea-quark momentum fractions at asymptotically large evolved scale.

The phenomenological success reported in Refs.\cite{Cui:2020dlm,Cui:2020tdf} and shortly described above strongly supports the conjecture about the IR agreement of $\widehat{\alpha}(k^2)$ and $\widetilde{\alpha}(k^2)$ within the IR. Then, assuming that an effective charge with an UV behavior featured by $\Lambda_{\text{QCD}}$=0.234 GeV is more suitable for phenomenological purposes, we argue that the running-interaction should be redefined just by replacing in Eq.\,\eqref{eq:dhatfromalpha} $\widehat{\alpha}(k^2)$ with $\widetilde{\alpha}(k^2)$ given by Eqs.\,(\ref{eq:alphaPIparam},\ref{eq:K2}), 
\begin{equation}\label{eq:dhatfromalphatilde}
\widehat{d}(k^2) = \widetilde{\alpha}(k^2) {\mathcal D}(k^2) \;;
\end{equation}
and it should be then applied to solving the gap equation with \eqref{eq:quark-prop2} and thus produce the quark mass function.    

\begin{figure}[ht!]
	\centering
	\includegraphics[width=0.95\columnwidth]{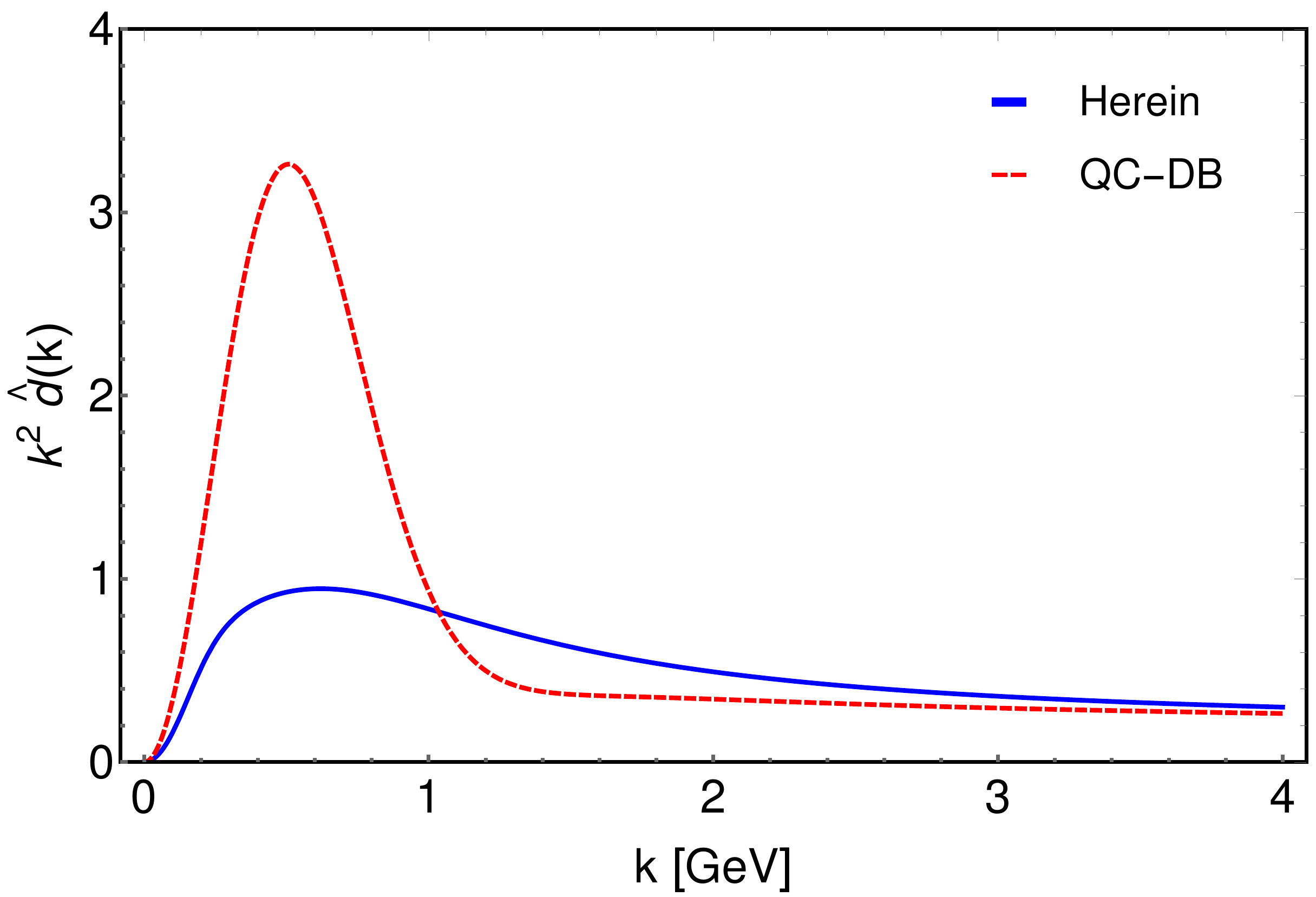}
	\caption{\textit{RGI interaction, $\hat{d}(k)$.} Effective running-interaction given by Eqs.\,(\ref{eq:alphaPIparam},\ref{eq:dhatfromalphatilde}) with the parameters  therein displayed (black solid line), compared to the  QC interaction (blue solid line), with typical model parameters for the so called DB kernel: $\omega=0.5$~GeV, $D\omega=(0.55~\text{GeV})^{3}$ and $\Lambda_{\text{QCD}}=0.234$~GeV\,\cite{Qin:2011dd,Chang:2013pq}. }
	\label{fig:dhat}
\end{figure}

In Fig.~\ref{fig:dhat}, we display the result for $\widehat{d}(k^2)$ given by  Eqs.\,(\ref{eq:alphaPIparam},\ref{eq:dhatfromalphatilde}) and, for comparative purposes, also include the well-known phenomenological model dubbed as Qin-Chang (QC) interaction~\cite{Qin:2011dd}. This QC model often comes along with the rainbow-ladder truncation of QCD's Dyson-Schwinger and Bethe-Salpeter equations~\cite{Chang:2013pq, Qin:2019hgk, Raya:2019dnh}, providing accordingly sensible results for the mass spectrum and structural properties of the vector and pseudoscalar mesons and ground-state baryons. The comparison made clearly apparent that, in using Eq.\,\eqref{eq:dhatfromalphatilde} for the running-interaction, the effective strenght of the QC interaction is reallocated in the QGV, enhanced herein by the ACM term given by Eq.\,\eqref{eq:ACMvertex}.

\section{The quark mass from Lattice QCD}

\begin{table}
\begin{tabular}{llcccr}
Symbol & $L^3\times T$  &a (fm)  &{$m_{\pi}$}(MeV) & $m_K$(MeV) & $N_{cfg}$ \\
\hline
64I  & $64^3\times128$& 0.0837(2) &139  &508  & 40  \\
48I  & $48^3\times 96$& 0.1141(2) &139  &499  & 40  \\
24I  & $24^3\times 64$& 0.1105(3) &340  &593  & 203  \\
24Ih  & $24^3\times 64$& 0.1105(3) &432  &626  & 143  \\
24Ih2  & $24^3\times 64$& 0.1105(3) &576  &660  & 85  \\
\hline
\end{tabular}
\caption{\label{table:lat} The parameters for the RBC/UKQCD 2+1 flavor configurations~\cite{Blum:2014tka,Mawhinney:2019cuc}: spatial/temporal size, lattice spacing,  pion mass with the degenerate light sea quark, kaon mass, and the number of configurations.}
\end{table} 

As it has been above stated, the main aim of this paper is confronting the quark mass function derived both from the DSE gap equation, capitalizing on the running-interaction featured by the effective charge given in Eq.\,\eqref{eq:alphaPIparam}, and from lQCD. For the latter, in order to obtain the mass function accurately, the use of a discretizated fermion action without addtional chiral symmetry breaking is essential. An optimal choice is the overlap fermion~\cite{Chiu:1998eu,Liu:2002qu} which satifies the Ginsburg-Wilson relation~\cite{Ginsparg:1981bj}. 

We have then employed the overlap fermion for the valence quark on five RBC/UKQCD 2+1 flavor Domain-Wall fermion ensembles with Iwasaki gauge action~\cite{Blum:2014tka,Mawhinney:2019cuc}, with their setup described in Tab.\,\ref{table:lat}. Two of these ensembles (labeled as 48I and 64I) are simulated with the physical light and strange quarks, to control the discretization error of the quark mass function; while the other three ensembles (24I/24Ih/24Ih2) with larger sea quark masses (and then heavier pion masses) are used to investigate the sea-quark mass dependence of the mass function\footnote{Note that the strange quark mass used by the three \emph{heavier} ensembles is $\sim$20\% larger than that on the physical-point ensembles, such a difference provides sensible hints on the impact of the strange quark mass on the mass function $M_f$}. 

 \begin{figure}[ht!]
	\centering
	\includegraphics[width=0.95\columnwidth]{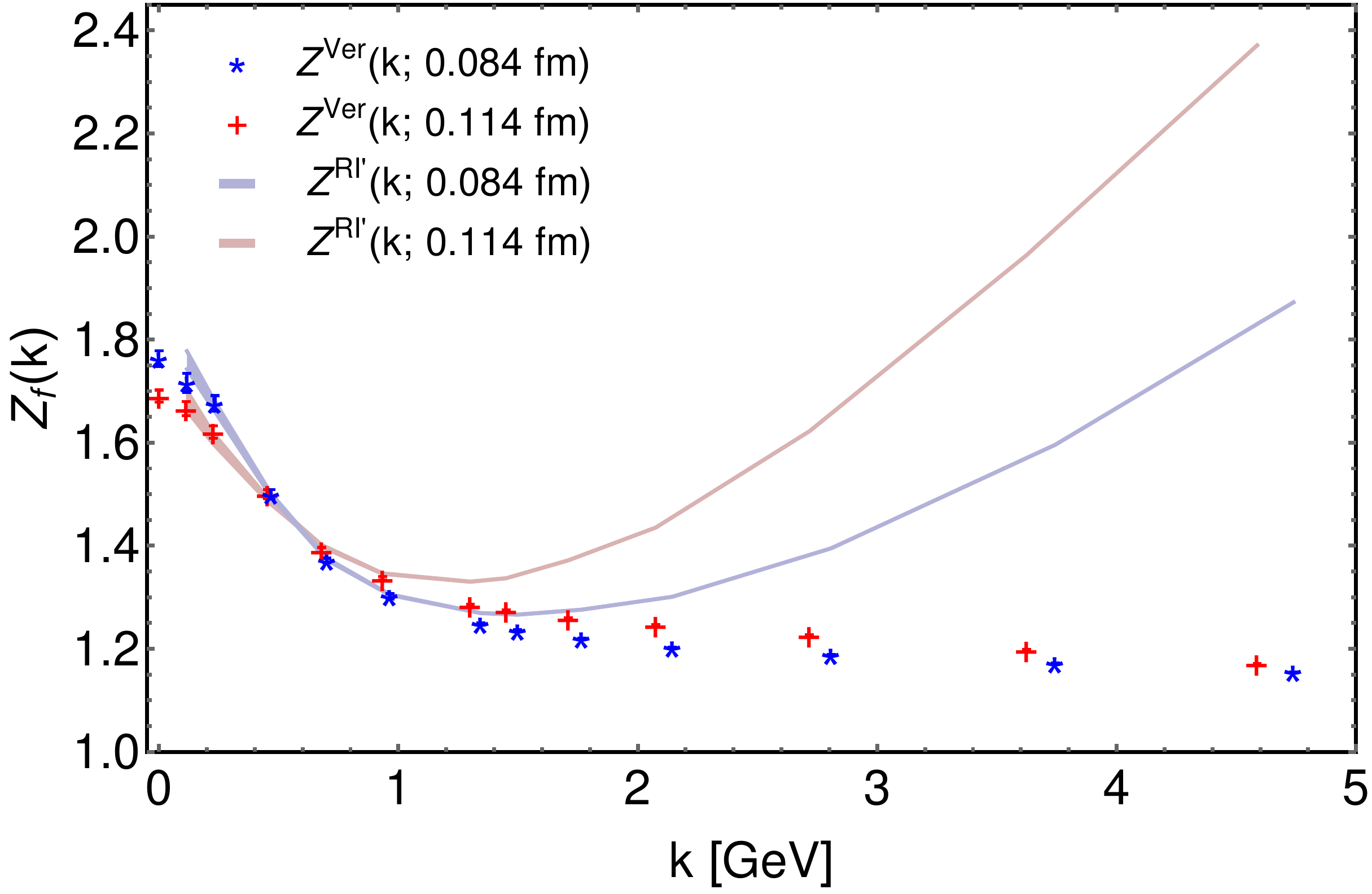}
	\includegraphics[width=0.95\columnwidth]{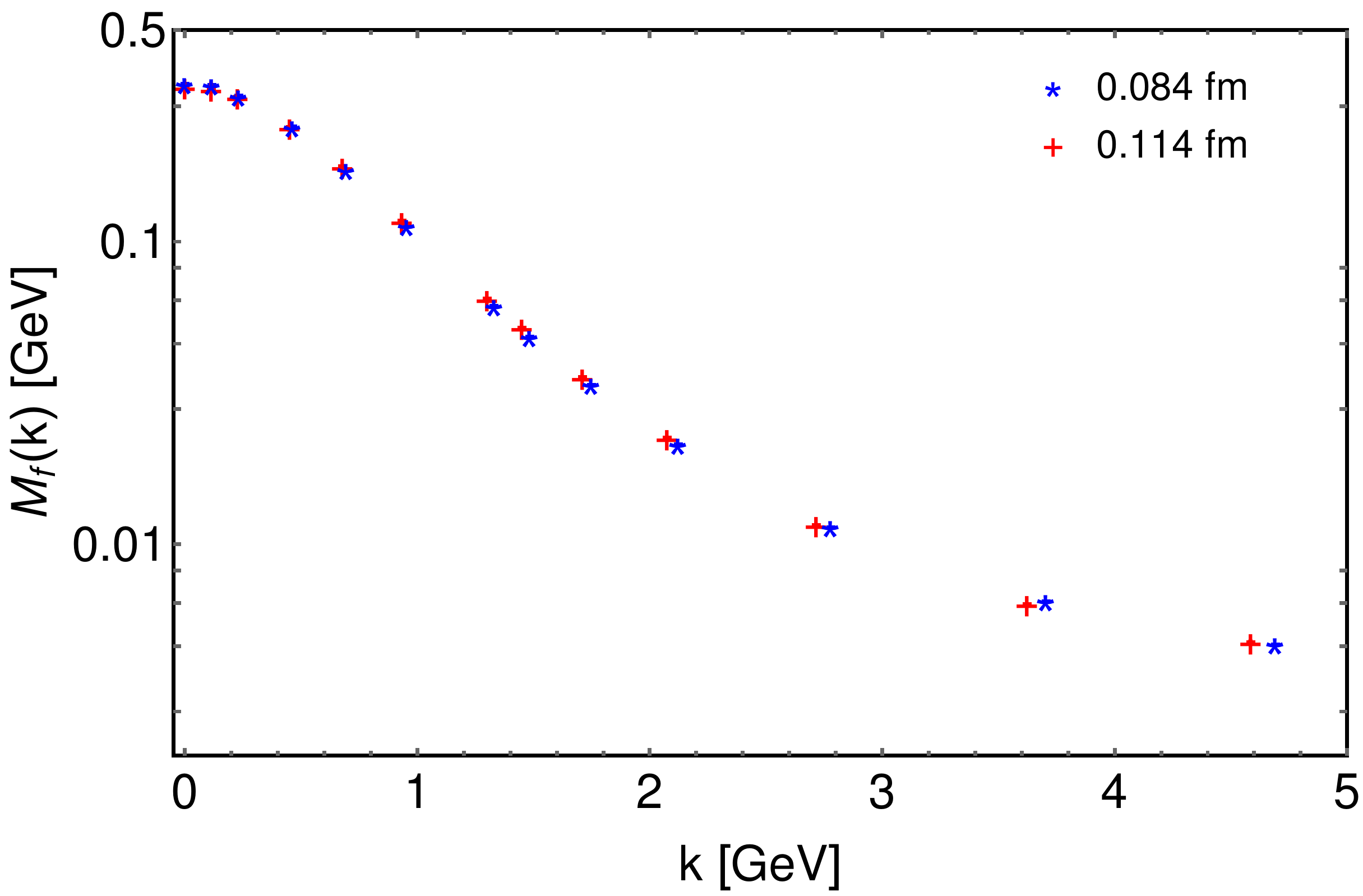}
	\caption{\textit{$Z_2$ and $M_k$ from Lattice QCD}. The $Z_2^{\textrm{Ver}}$ using the vector vertex correction defintion Eq.~(\ref{eq:ri2}) at two lattice spacings (blue and purple data points) are quite close to each other; But $Z_2^{\textrm{RI'}}$ from the quark propagator definition Eq.~(\ref{eq:ri1}) suffer from obious discretization error while approch to $Z_2^{\textrm{Ver}}$ in the contiuum limit (upper panel). The consistency of $M^{\textrm{Ver}}_k$ at two lattice spacings shows that the discretization error is controlled with the vertex definition (lower panel).}
	\label{fig:self_energy}
\end{figure}

The quark mass function can be computed from these lattice QCD ensembles as
\begin{subequations}
\label{eq:ri1}
\begin{eqnarray}
\label{eq:MRI}
M^{\textrm{RI'}}_{f}(p^2)&=&\frac{1}{12}\text{Tr}[S^{-1}_{\textrm{lat}}(p)]/Z^{\textrm{RI'}}_2(p^2)\;,\\
Z^{\textrm{RI'}}_{2}(p^2)&=&\frac{1}{12}\text{Tr}[p\!\!\!/\,S^{-1}_{\textrm{lat}}(p)]/p^2\;;
\end{eqnarray}
\end{subequations}
with $S_{\textrm{lat}}(p)\equiv\sum_{x,y} e^{-ipx}S(p,x)/V$ [$V$ standing for the 4D volume] defined in terms of the Landau-gauge propagator at a fixed volume source, $S_{\textrm{lat}}(p,w)= \langle  \psi(w) \sum_{y}{\bar{\psi}(y)e^{ipy}}\rangle$ (the statistical uncertainty is then reduced by a factor $\sqrt{V}$, thus allowing for a very precise result at low $p^2$). The superscript RI' stands for the modifed regularization independence scheme.  

However, $Z_2^{\textrm{RI'}}$ remains not well-defined at $p^2=0$ and prevents therewith from a direct extraction of $M_f(0)$. One can alternatively define from the transverse part of the vertex correction,
\begin{equation}
\label{eq:ri2}
Z^{\textrm{Ver}}_2(p^2)=\frac{Z_V}{36}\textrm{Tr}[\gamma_{\nu}\Lambda(p, \gamma_\mu)T_{\mu\nu}(p)] \;,  
\end{equation}
where $Z_V=\langle \pi|\bar{\psi}\gamma_4\psi|\pi\rangle/\langle \pi|\pi\rangle$ is defined through the corresponding hadron matrix elements and 
\begin{subequations}
\begin{eqnarray}
\Lambda(p, \Gamma)&=&\frac{S_{\textrm{lat}}^{-1}(p) \left\langle \Gamma \right\rangle   S_{\textrm{lat}}^{-1}(p)}{V}\;,
\\
\left\langle \Gamma \right\rangle &=& \left\langle \sum_w {\gamma_5}S_{\textrm{lat}} ^{\dagger}(p,w){\gamma_5}\Gamma S_{\textrm{lat}}(p,w) \right\rangle \;;
\end{eqnarray}
\end{subequations}
$Z^{\textrm{RI'}}_2$ and $Z^{\textrm{Ver}}_2$ being exactly the same under dimensional regularzation~\cite{Gracey:2003yr}, the latter defined at $p^2=0$ without sigularity. Another ace of this vertex definition of $Z^{\textrm{Ver}}_2$ is its being less affected by discretization errors\footnote{This can be well understood as $Z^{\textrm{RI'}}_2$ is defined through the quark proapgator $S_{\textrm{lat}}$, affected by large discretization errors which become majorly canceled by its inverse in the vertex defintion $Z^{Ver}_2$.}, as shown in the upper panel of Fig.~\ref{fig:self_energy}. 


Thus, one can define $M^{\textrm{Ver}}_{f}(p^2)$ as in Eq.\,\eqref{eq:MRI} but replacing $Z_2^{\textrm{RI'}}$ with $Z_2^{\textrm{Ver}}$ given by \eqref{eq:ri2}. Results for $M^{\textrm{Ver}}_{f}(p^2)$ obtained from the two lattice ensembles at the physical point and different lattice spacings are displayed in the lower panel of Fig.~\ref{fig:self_energy}, and are shown to be very consistent, thus confirming that discretization errors are systematically under control. Consequently, we will use $M^{\textrm{Ver}}_{f}(p^2)$ for the lQCD estimates of the quark mass function, but will omit the superscript in the following. 

 \begin{figure}[t!]
	\centering
	\includegraphics[width=0.95\columnwidth]{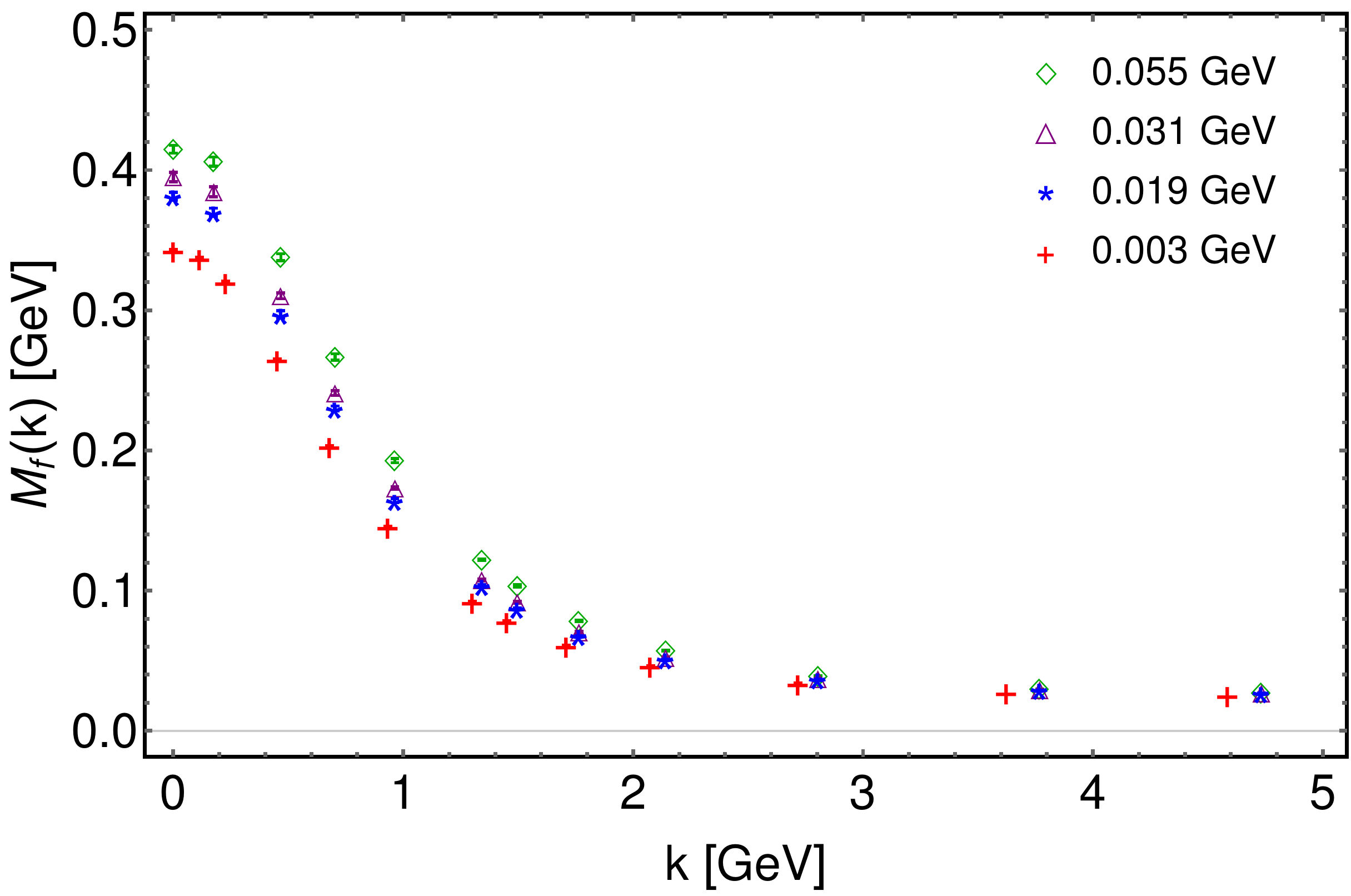}
	\caption{The quark mass function obtained as explained in the text, with $Z_2^{\textrm{Ver}}$ given by \eqref{eq:ri2}, from the physical-point ensembles and from the three others with larger sea-quark masses (Tab.\,\ref{table:lat}). A sea-quark mass effect is clearly made apparent at small $k$.}
	\label{fig:Mass_sea}
\end{figure}

On the other hand, the physical-point results of the mass function appear compared in Fig.\,\ref{fig:Mass_sea} to the same computed from the three ensembles with heavier sea-quark (and pion) but the same valence-quark mass. The sea-quark mass is clearly seen to have a sizeable impact at low momenta and this therefore implies the need of approaching the \emph{unitary} point, at which valence- and sea-quark masses are the same, aiming at a comparison with the gap-equation results. 

To this purpose, the multi-mass algorithm has been applied with the overlap fermions to produce mass function results for ten different valence-quark masses on all the three ensembles with heavier pion masses in Tab.\,\ref{table:lat}. Then, we have made interpolations from the results with the three ensembles and have therefrom produced mass functions observing the unitarity condition and four different current quark masses roughly ranging from 2 to 60 MeV. This is the output which the gap-equation results are to be confronted to in the next section.

\section{The quark mass function from the gap equation}

Then, the gap equation \,\eqref{eq:gapEqs} reshaped in Eq.\,\eqref{eq:quark-prop2} is to be solved, as described in Secs.\,\ref{sec:gapEq} and\;\ref{sec:eff-charge}, applying the running-interaction given by Eqs.\,(\ref{eq:alphaPIparam},\ref{eq:dhatfromalphatilde}) and the QGV defined through Eqs.\,(\ref{eq:vertex},\ref{eq:ACMvertex}). The former becomes completely defined by the effective charge, the PT-BFM truncation of DSEs in the QCD gauge sector and the lattice inputs for the two-point Green's functions. The latter is expected to capture efficaciously all the relevant nonperturbative dynamics and, if so, the two only free parameters $m_0$ and $\eta$ should serve for any flavor. In practice, this means that, by tuning these two parameters, the quark mass functions resulting from lQCD and from solving the gap equation should consistently agree with each other for any fixed value of the current quark mass. In particular, for those introduced in the last section. 

 \begin{figure}[t!]
	\centering
	\includegraphics[width=0.95\columnwidth]{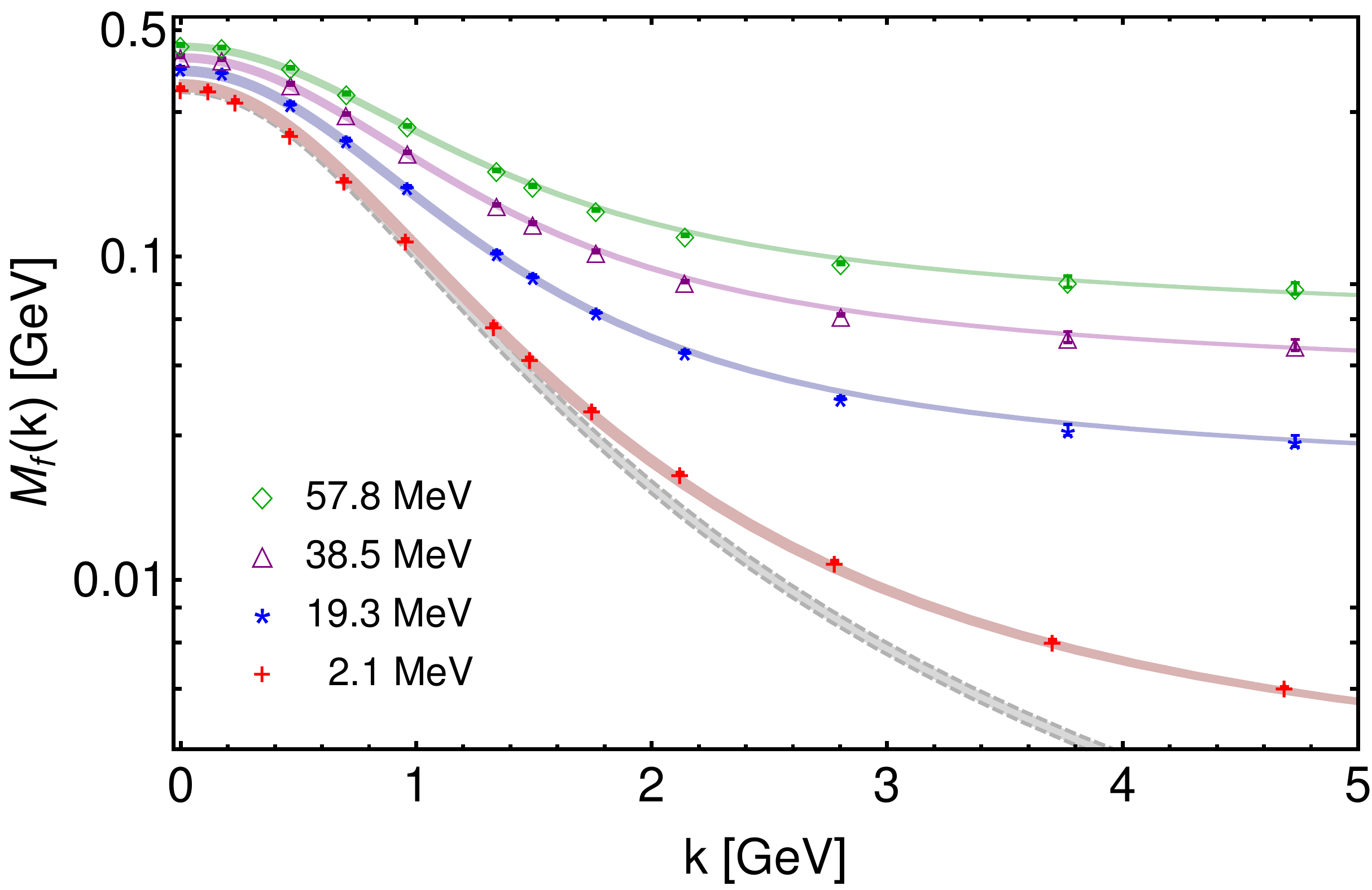}
	\caption{\textit{Mass functions.} lQCD mass functions (data points) and DSE results from the Ansatz in Eqs.~(\ref{eq:quark-prop2}-\ref{eq:ACMvertex}). The current quark masses are (in MeV): 57.8, 38.5, 19.3, 2.1 and the chiral limit (top to bottom). The gray band at the bottom corresponds to the chiral limit. The band in the gap equation results accounts for the variation of $\eta\in[1.27,\;1.32]$. Statistical errors for the lattice data and uncertainties rooting on the interpolation applied to reach the unitary point are of the order of 1 \% and have not been thus displayed, given the comparative purpose of the plot and its accurateness.}
	\label{fig:BCtau5}
\end{figure}

Indeed, as it is shown in Fig.\,\ref{fig:BCtau5}, after specializing $m_0=2$ GeV and\footnote{Although the running-interaction and the QGV are both expected to be flavor independent, and hence independent on the current-quark mass, the interaction is affected by the mass of the light sea-quark. Then, as the masses of one or another are tuned to be the same in the unitary point through interpolation, we account for the latter by attributing a small range of uncertainty to $\eta$.} $\eta\in[1.27,\;1.32]$, and applying the same current quark masses, the agreement of lQCD and gap equation results is strikingly good: the gap equation solutions accurately reproduce the lQCD estimates for all the considered quark masses, roughly ranging from 2 to 60 MeV. This success can be interpreted as a strong indication in favour of the effective charge approach to solve the gap equation for phenomenological applications.  
 
The latter in mind, we have also solved the gap equation in the chiral limit (the gray band in Fig.\,\ref{fig:BCtau5}) and then applied the formula derived in~\cite{Roberts:1994hh,Cahill:1985mh},
    \begin{eqnarray}
        \nonumber
        (f_\pi^{(0)})^2 &=& \frac{3}{8\pi^2} \int dp^2\; p^2 B^2(p^2)\Big(\sigma_v^2 -2 [\sigma_s \sigma_s'+p^2 \sigma_v \sigma_v'] \\
        \label{eq:fpi2}
        &-& p^2[\sigma_s \sigma_s'' - \sigma_s' \sigma_s']-p^4 [\sigma_v \sigma_v'' - \sigma_v' \sigma_v'] \Big)\;,
    \end{eqnarray}
where $B(p^2)$, defined from Eq.\,\eqref{eq:ACMvertex}, is here specialized for vanishing current quark mass (the superindex `$(0)$' denotes this particular limit); and the functions $\sigma_{s,v}=\sigma_{s,v}(p^2)$ are simply quark propagator dressing functions, 
\begin{eqnarray}
S(p) &=& -  i \gamma \cdot p\;\sigma_v(p^{2}) + \sigma_s(p^{2}) \;;
\end{eqnarray}
thus reading
\begin{eqnarray}
\sigma_s(p^2)&=&M(p^2)\sigma_v(p^2) \;, \nonumber \\
\sigma_v(p^2)&=&\frac{Z(p^2)}{p^2+M^2(p^2)}\;,
\end{eqnarray}
in terms of the quark mass function $M(p^2)$ and the dressing function $Z(p^2)$ from Eq.\,\eqref{eq:gapEqs} in the chiral limit. We have therewith obtained $f_\pi^{(0)}=89.5(1.8)$ MeV, where the error accounts for the variation of $\eta$. This result is fairly consistent with the lQCD average at the $N_f=2$ chiral limit, 86.2(5) MeV~\cite{Aoki:2019cca}, both agreeing within an error uncertainty of two $\sigma$'s. In the chiral limit, the renormalization point dependent chiral quark condensate (CQC) can be expressed as follows:
\begin{equation}
    \label{eq:condensate}
    -\textless\bar{q}q\textgreater_\zeta^{(0)} = Z_4 N_c \mbox{Tr} \int_q^\Lambda S(q;\zeta) \;,
\end{equation}
where $Z_4$ is the mass term renormalization constant, such that $Z_4 m_f^\zeta = Z_2 m_f^{\text{bm}}$. The CQC is often considered as an order parameter of dynamical chiral symmetry breaking~\cite{Sultan:2018qpx}, since its non-zero value appears only non-perturbatively. It moreover coincides with the $N_f=2$ chiral limit value of the so called in-pion condensate~\cite{Brodsky:2010xf,Chang:2011mu} and, at $\zeta=19$ GeV, amounts to  $-\textless \bar{q}q \textgreater_\zeta^{(0)} = (0.283(4) \; \textrm{MeV})^3$ from the quark propagators herein obtained for $\eta\in [1.27,\;1.32]$. We eventually appeal to the Gell-Mann-Oakes-Renner formula~\cite{GellMann:1968rz}:
\begin{equation}
    m_{\pi^\pm}^2 = -m_{ud}^\zeta\frac{\textless \bar{q}q \textgreater_\zeta^{(0)}}{(f_\pi^{(0)})^2}+\mathcal{O}(m_{ud}^2)\;,
\end{equation}
where $m_{ud}:=m_u^\zeta + m_d^\zeta \approx 0$. In order to be left with $m_{\pi^\pm}=0.140$ GeV, the obtained CQC and $f_\pi^{(0)}$ demand $m_{ud}^\zeta\simeq 6.8$ MeV, which is in excellent agreement with the empirical value~\cite{Zyla:2020zbs}. This result encourages the perspective of an accurate extraction of $m_\pi$ from a fully symmetry preserving approach to solve the meson Bethe-Salpeter equation. It is wothwhile hilighting that the CQC and quark mass values herein derived are not under the MS-bar scheme, and a comparison with lQCD averages~\cite{Aoki:2019cca} requires thus further investigation.

\section{conclusion}

A gap-equation running-interaction has been defined on the ground of a phenomenological effective charge, which in its turn shares the IR behaviour, and particularly the vanishing-momentum saturation, with the process-independent effective charge defined within the framework of the PT-BFM truncation of gauge-field DSEs. Obtaining therewith the quark mass function from the gap equation, we have bridged from the QCD gauge sector, where key nonperturbative features endows the gluon with a dynamical running mass, to the matter sector where the emergence of hadron masses takes place. This connection can be seen to inspire a deep understanding of the origin of most of the visible mass in the Universe.

Aiming at a support paving the way for this connection of gauge and matter sectors, results for the quark mass function from the gap equation and lattice QCD have been successfully confronted. In solving the gap equation, the fully quark-gluon vertex became enhanced by an Anomalous Chromomagnetic Moment term, tuned by fixing two flavour-independent parameters which, amounting to sensible values, made possible an accurate description of the lattice results. For the latter, Domain-Wall and overlap fermion actions have been respectively applied for the sea and valence quarks, guaranteeing thus the best chiral properties on the lattice. Five different lattice sets of configurations simulating dynamical quarks with different masses, including two at the physical point with different lattice spacings, have been used to produce quark propagators with ten different valence-quark masses each. We have then combined the results to extract and deliver the quark mass function for four different current-quark masses, the same ones that have been implemented for the gap equation.  

Finally, after succeeding with a congruent comparison of lattice and gap-equation results for the quark mass functions, we have computed the pion decay constant and quark condensate in the $N_f=2$ chiral limit; and have accommodated their values into the Gell-Mann-Oakes-Renner formula with empirically consistent values for the pion and light quark masses.

\section*{Acknowledgement}
We are grateful for constructive comments from Craig D. Roberts, and we thank $\chi$QCD collaboration for providing us their overlap RI/MOM renormalization constants on the RBC/UKQCD configurations. The numerical lQCD calculation is performed using the GWU-code~\cite{Alexandru:2011ee,Alexandru:2011sc} through HIP programming model~\cite{Bi:2020wpt}, and supported by Strategic Priority Research Program of Chinese Academy of Sciences, Grant No. XDC01040100, and also HPC Cluster of ITP-CAS. Y-B. L is supported by the National Natural Science Foundation of China under Contract No. 11875169. Y-B. Y. is supported in part by the Strategic Priority Research Program of Chinese Academy of Sciences, Grant No. XDC01040100 and  XDB34030303, and a NSFC-DFG joint grant under grant No. 12061131006 and SCHA 458/22; J.R.Q. is supported by the Spanish MICINN grant PID2019-107844-GB-C2, and the regional Andalusian project P18-FR-5057.

\bibliography{main}

\end{document}